\documentclass[aps,pra,preprint,showpacs,preprintnumbers,amsmath,amssymb,footinbib]{revtex4-1}
\usepackage{graphicx}
\usepackage{bm}
\usepackage{amsmath,amssymb}
\usepackage[colorlinks=true, pdfstartview=FitV, linkcolor=red,citecolor=blue, urlcolor=blue]{hyperref}
\usepackage{color}
\usepackage[final]{changes}
\usepackage{simplewick}

\newcommand{\Tr}[1] { {\rm tr} \left[ #1 \right]}

\begin{document}

\preprint{RBRC-?}

\title{Correlation functions of Polyakov loops at tree level}

\author{Robert D. Pisarski}
\email{pisarski@bnl.gov}
\affiliation{Department of Physics,
Brookhaven National Laboratory, Upton, NY 11973, USA}
\affiliation{RIKEN/BNL Research Center,
Brookhaven National Laboratory, Upton, NY 11973, USA}
\author{Vladimir V. Skokov}
\email{vladimir.skokov@wmich.edu}
\affiliation{Department of Physics,
Western Michigan University, 1903 W. Michigan Avenue, Kalamazoo, MI 49008}

\begin{abstract}
We compute the correlation functions of Polyakov loops
in $SU(N_c)$ gauge theories by explicitly summing
all diagrams at tree level in two special cases,
for $N_c = 2$ and $N_c = \infty$.  
When $N_c =2$ we find the expected
we find Coulomb-like behavior at short distances, 
$\sim 1/x$ as the distance $x \rightarrow 0$.
In the planar limit at $N_c = \infty$ 
we find a weaker singularity, $\sim 1/\sqrt{x}$ as $x \rightarrow 0$.
In each case, 
at short distances the behavior of the correlation functions between
two Polyakov loops, and the corresponding Wilson loop, are the
same.
We suggest that such non-Coulombic behavior is an artifact of the planar limit.
\end{abstract}
\maketitle

\section{Introduction}

In gauge theories at a nonzero temperature $T$, 
the order parameter for
deconfinement in a $SU(N_c)$ gauge theory is the Polyakov loop,
\begin{equation}
L (x) = \frac{1}{N_c} \; {\rm tr}\; {\cal P} 
\exp \left( i g \int_0^{1/T} A^0(x,\tau) d \tau \right) \; .
\label{Eq:L}
\end{equation}
This loop is related to the propagator of an infinitely massive test quark,
which sits at a spatial point $x$, and just propagates forward in
imaginary time, $\tau: 0 \rightarrow 1/T$.  
A quantity invariant under small gauge transformations is obtained by
tracing over color, denoted by $\rm tr$;
${\cal P}$ represents path ordering in $\tau$, 
$g$ is the gauge coupling, and $A^0$
the vector potential in the $\tau$ direction.  

To be definite, we consider 
the case of a test particle, and its associated loop, in the fundamental
representation of a $SU(N_c)$ gauge theory.  (As for test particles, though,
there are (independent) loops for all irreducible representations.)
For two colors the loop is real, but for three or more colors the anti-loop,
$L^*(x)$, is an independent quantity, representing the 
propagation of an anti-particle backwards in (imaginary) time.

The correlation functions of Polyakov loops are fundamental
to our understanding of the behavior of gauge theories in a thermal bath.
These have been studied in great detail, both perturbatively, and
from numerical simulations on the lattice
\cite{Gava:1981qd, Arnold:1995bh, Burnier:2009bk, Brambilla:2010xn, Lo:2013etb,*Lo:2013hla}.  
In general, understanding the behavior of the loop correlation functions
is involved, requiring the use of effective Lagrangians and non-perturbative
input, such as from numerical simulations on the lattice
\cite{Borsanyi:2015yka}.

This is not true in an Abelian gauge theory, where at least at tree level,
one can immediately compute the correlation
function between two loops.  In this case, the path need not be straight
but can be arbitrary.  A source $J^\mu$ couples to the photon as
$\sim \exp(i \oint ds \, J^\mu A_\mu)$, where ``$s$'' the worldline of the
test particle.  It is then trivial to integral over over $A_\mu$,
\begin{equation}
\sim \exp \left(- \frac{e^2}{2} \oint ds \oint ds' \; 
J^\mu \Delta_{\mu \nu} J^\nu \right) \; ,
\label{abelian}
\end{equation}
and obtain the sources tied together by a 
photon propagator, $\Delta_{\mu \nu}$.

This exercise is not possible 
in a non-Abelian gauge theory, where the fields are matrices
and do not commute.  As an exercise, in this paper
we compute the correlation functions
between Polyakov loops at tree
level.  Even in this instance the generalization of Eq. (\ref{abelian}) is
not possible for arbitrary $N_c$.  This is basically because for
$3 \leq N_c < \infty$, one cannot determine the element of the Lie group,
the Wilson line, 
from that of the Lie algebra, $A^0$.
We compute the two point functions of loops
in two special limits, for two colors, and in the planar limit,
$N_c \rightarrow \infty$ with $g^2 N_c$ held fixed.  

As a computation at tree level, our results are valid only in the limit
when the temperature $T$ is very large, so that the running coupling
constant, $g^2(T)$, is small.  We compute the correlation functions
between two loops at an arbitrary distance, $x$; this will always
appears with the only dimensional scale in the problem, the temperature
$T$, as $xT$.  At large distances,
this must agree with the results of an ordinary perturbative expansion
computed to tree level.
Because each loop is a trace over color indices, the leading diagram, from
one loop gluon exchange, cancels, and so unlike
the Abelian theory, there is no term $\sim g^2/(xT)$.
The leading diagram is given by 
two gluon exchange, which is $\sim g^4/(xT)^2$.  
This is valid at both $N_c = 2$ and $N_c = \infty$, and is a trivial
check on our results.

The natural expansion parameter in the correlation functions 
is $\sim g^2/(x T)$.  Thus at short distances, naive perturbation theory
cannot be used.  When the loops are very close together, though,
$x T \ll 1$, though, physical intuition suggests
that it shouldn't matter how the loops are tied off from one another.
Thus the result should be the same as for a Wilson loop that is very long.
As a Wilson loop is a single trace over color, single gluon exchange 
contributes, so the result at small distances should be Coulombic, 
$\sim g^2/(x T)$.  

We find that at short distances, in both cases the leading behavior at
short distances is the same for the correlation between a Polyakov loop
and an anti-loop (or two Polyakov loops), and the corresponding Wilson loop.
For two colors, at short distances the behavior is Coulomb-like, although
the coefficient is not identical to that from single gluon exchange.
For an infinite number of colors, the behavior is milder,
$\sim \sqrt{g^2/(xT)}$.  
Although our results are only valid to tree level, for arbitrary distances
we obtain non-trivial functions of $g^2/(xT)$ which can be written
in closed form.

In the Conclusions, Sec. (\ref{sec:Conclusions}),
we discuss why the non-Coulombic behavior 
at short distances is an artifact of the planar
limit, and discuss how to compute the short distance behavior
from effective theories \cite{Burnier:2009bk, Brambilla:2010xn}.
Our study was inspired by recent studies of the susceptibilities of Polyakov
loops by Lo, Friman, Kaczmarek, Redlich, and Sasaki 
\cite{Lo:2013etb,*Lo:2013hla}.  
We also discuss in the Conclusions the implications of our analysis for
the measurements of such loop susceptibilities.

\section{Conventions}

Since we only compute Polyakov and Wilson loops, at the outset it is
convenient to adopt static gauge, $\partial_0 A^0(\tau,\vec{x}) = 0$.  It
is also necessary to fix the gauge dependence of spatial gluons,
$A^i$, at a given time, but at tree level the spatial gluons do not
enter into the correlators of Polyakov loops or of the Wilson loops which
we consider.

The advantage of static gauge is that for the Polyakov loop,
time ordering can be ignored, with 
the Polyakov loop just the exponential of a single field
$A^0(\vec{x})$.  For single gluon exchange, it is clear
that in an arbitrary gauge, time ordering
gives the same result as in static gauge.  It is not so obvious
for the exchange of two or more gluons, but in
the end, both Polyakov (and Wilson) loops are gauge invariant.  
In any case, what we are really interested in is how the 
correlator of the exponential of non-Abelian $A^0$ fields is related to the
correlator of the $A^0$ fields.

The gluon field $A^0 = A^0_a t^a$, normalizing the
generators as ${\rm tr}( t^a t^b) = \delta^{a b}/2$,
$a,b = 1\ldots N_c^2-1$.
For any $N_c$, in static gauge
the gluon propagator for $A^0_a$ is color diagonal,
\begin{equation}
\Delta^{a b}_{0 0}(\vec{x}) = \langle A^0_a(\vec{x}) A^0_b(0) \rangle
= \delta^{a b} \, T \, \Delta(\vec{x}) \;\;\; , \;\;\;
\frac{1}{T} \; \Delta(\vec{x}) = 
\frac{1}{T} \; \int \frac{d^3 k}{(2 \pi)^3} \; 
\frac{{\rm e}^{ i \, \vec{k} \cdot \vec{x} }}{\vec{k}^{\, 2}}
= \frac{1}{4 \pi}\; \frac{1}{ x\, T} \; .
\label{gluon_prop}
\end{equation}
Since we only compute two point functions, which are a function of a single 
spatial distance $\vec{x}$, henceforth we replace $\vec{x}$ by
$x = |\vec{x}|$.  We introduce $\Delta(x)/T$
since Polyakov loops are dimensionless, and so at tree level their 
correlation functions are functions only of 
$x T$.  

As our computation is only at tree level,
several well known physical effects do
not enter.  Beyond leading order, the Debye mass, $\sim g T$,
arises to screen
correlation functions involving $A^0$.  Also, the renormalization mass
scale enters to ensure the coupling constant runs.  

The virtue of computing at tree level is that it is 
mathematically well defined.
Of course then one must be careful to consider the limitations of this
approximation, as we discuss in the Conclusions, Sec. (\ref{sec:Conclusions}).

\section{Two colors}

For two colors, we can use the well known property of Pauli matrices,
$\sigma^a$, 
and evaluate any element of the Lie group from that of the Lie algebra,
\begin{equation}
\exp\left( i \, \beta^a \, \sigma^a \right) =  
\cos\left( \beta \right) 
+ i \; \widehat{\beta}^a \, \sigma^a \; 
\sin\left( \beta \right) \; ,
\label{two_colors_expansion}
\end{equation}
where $\beta^a = \widehat{\beta}^a \beta$, $(\widehat{\beta})^2 = 1$.

The vector potential is $A^0 =
A^0_a \sigma^a/2$.  Since
$A^0_a$ are the only degrees of freedom which enter, we denote it
simply as $A_a$, and use a convention for repeated indices in color,
\begin{equation}
A_a^2 = \sum_{a = 1}^{3} (A_a)^2 \; .
\end{equation}
Hence the Polyakov loop is given by the first term, from the cosine:
\begin{equation}
L(x) = \frac{1}{2} \; \Tr{ 
e^{i g A^0_a(x) \sigma^a/(2 T)} } =  
\cos\left( \frac{g}{2 T} \sqrt{A_a(x)^2} \right) =
\sum_{n=0}^\infty   \frac{(-1)^n}{(2n)!} 
\left( \frac{g^2}{4 T^2} \; A_a(x)^2 \right)^n 
\label{Eq:two_color_identity}
\end{equation}
This also shows that for two colors the loop is always
real, $L = L^*$.  The correlation function of two loops is then
\begin{equation}
\langle L(x) L(0) \rangle  = 
\left\langle \sum_{n=1}^\infty   \frac{(-1)^n}{(2n)!} 
\left( \frac{g^2 }{4 T^2} A_a(x)^2\right)^n \;
\sum_{n'=1}^\infty   
\frac{(-1)^{n'}}{(2n')!} 
\left( \frac{g^2}{4 T^2} A_b(0)^2\right)^{n'} \right\rangle 
\label{Eq:two_one}
\end{equation}
We ignore self energy corrections, which vanish with 
dimensional regularization.  Then the correlator only receives
contributions from terms where $n=n'$,
\begin{equation}
\label{LLd}
\left\langle L(x) L(0) \right\rangle  = 
\sum_{n=0}^\infty \; \frac{1}{((2n)!)^2} \; 
\left( \frac{g}{2 \, T} \right)^{4 n}
\left\langle  \left(A_a(x)^2\right)^n
\left( A_a(0)^2 \right)^n \right\rangle \; .
\end{equation}

If only one color contributed, then this correlation function would 
follow immediately.  Because for two colors there are three gluons,
we have to solve the problem of how many ways to tie three fields together
in all possible ways.  We do this in two ways, as a cross check on each method.
The first method is combinatoric, just a matter of counting different
fields together.
The second is analytic, and may be useful for other problems,
when counting permutations might be confusing.

\subsection{Combinatoric analysis}

We solve a more general problem, that of computing the combinatoric factor
in Eq. (\ref{LLd}) when the sum over fields is from $a = 1$ to $m$, instead
of $m = 3$.  We work inductively, starting from $m=1$, where the answer
is immediate, and then generalize to arbitrary $m$ by induction.

The problem is to determine the number of ways in which we can tie
the fields
\begin{equation}
\left\langle \left(A_a(x)^2 \right)^n  \;\;
\left(A_b(0)^2 \right)^n \right\rangle  \; 
\end{equation}
together.  We denote this as ${\cal D}(2n,m)$.

For a single component field, $m=1$, the answer is obvious.
The first $A(x)$ can connect with any of $2n$ $A(0)$'s;
the second $A(x)$ can connect to any of the remaining $2n -1$ $A(0)$'s,
and so on.  This gives the usual factorial, 
${\cal D}(2n,1) = (2n)(2n-1)\ldots = (2 n)!$.

To solve for $m > 1$, we start with some special cases.

When $n = 1$, the first $A_a(x)$ can connect to one of the two 
$A_b(0)$'s.
The second $A_a(x)$ must then connect to the other
$A_b(0)$.  Since $a=b$, the sum over $a$ gives $m$, and so
\begin{equation}
{\cal D}(2,m) = 2! \; m \; .
\label{Eq:first_ex}
\end{equation}

Now consider $n=2$. The first $A_a(x)$ can connect to any of the four 
$A_b(0)$'s.  There are then two possibilities.  
The second $A_a(x)$
can connect to the partner of the first $A_b(0)$, which gives a factor
of $m$.  Tying the second pair together gives ${\cal N}(1,m) = 2 m$,
or $8 m^2$ in all.

Alternately, the second $A_a(x)$ can
tie to a different pair, $A_c(0)^2$.
It can do so in one of two ways.
This leaves two ways for the third $A_d(x)$ to connect to
one of the two remaining $A(0)$'s.  All four
$A$'s have the same index, so there is one sum over $a$, or
$16 m$ in all.  The sum is $8 m^2 + 16 m$, or
\begin{equation}
{\cal D}(2,m) = 4!! \; (m+2)m \; .
\label{Eq:second_ex}
\end{equation}
where $n!! = n(n-2)\ldots\; $.

For any $n$, the numbers of ways to connect with the
highest powers of $m$ is evident, just $m^n$.  The first $A_a(x)$ can connect
with any of the $2n$ $A_b(0)$'s.  The second $A_a(x)$ has to 
connect with the pair of $A_b(0)$, which gives $2n \times \; m$.  
The third $A_c(x)$ can connect to any of the remaining $(2n-2)$
$A_d(0)$'s, and so on.  Thus for large $m$, the leading term is
\begin{equation}
{\cal D}(2n,m\rightarrow \infty) 
= (2n) \; m \; (2n-2) \; m \ldots 2 \; m =
(2n)!!  \; m^n \; .
\end{equation}

Looking at the results for $n=1$ and $2$, and the result for large
$m$, a guess for arbitrary $m$ and $n$ is
\begin{equation}
{\cal D}(2n,m) = (2n)!! \; \frac{(2n-2+m)!!}{(m-2)!!} \; .
\label{final}
\end{equation}
When $m = 1$ we define $(-1)!! = 1$, so 
${\cal D}(2n,1) = (2n)!! (2n-1)!! = (2n)!$.

To establish the result for $m > 1$ we use induction,
and consider how
${\cal D}(2n+2,m)$ is related to ${\cal D}(2n,m)$.
As always, the first $A_a(x)$ can connect to $2n+2$ different
$A_b(0)$'s.
There are two ways in which the second $A_a(x)$ can connect
to a field $A_b(0)$.  The simplest 
is if it connects to the pair of the $A_b(0)$ which the first $A_a(x)$
connected to.  There is only one such 
$A_b(0)$ to connect to.  The sum over $a=b$ gives a factor of $m$,
so we have $(2n+2)m$ times the 
number of ways of tying $n$ pairs of $A_a$ 
together, which is ${\cal D}(2n,m)$.

Alternately, the second $A_a(x)$ can tie with a
$A_{c}(0)$, which is in a {\it different} pair than the first $A_a(x)$
connected to.  There are 
$2n$ different $A_c(0)$'s to choose from.  
This possibility looks complicated at first, but explicitly what we have is
\begin{equation}
\left\langle A_a(x) A_b(0) \right\rangle \; 
\left\langle A_a(x) A_c(0) \right\rangle \;
\left\langle \left(A_c(x)^2\right)^n \; 
A_b(0) A_c(0) \;
\left(A_d(0)^2\right)^{n-1}\right\rangle \; .
\label{Eq:comb1}
\end{equation}
Now $A_b(0)$ and
$A_c(0)$ start out as being in different pairs.
However, the propagators are diagonal in the color indices,
and evaluating the propagators
in Eq. (\ref{Eq:comb1}) gives $a=b$ and $b=c$.
After doing so, we have that $A_b(0)$ and $A_c(0)$ have
the same indices.  There is no factor of $m$, because we still
have to tie $A_b(0)$ and $A_c(0)$ to other $A(x)$'s.
Consequently, we are left with $(2n+2)(2n)$ times 
the number of ways of tying $n$ pairs together, which is ${\cal D}(2n,m)$.

The sum of these two terms is
\begin{equation}
{\cal D}(2n+2,m) = (2n+2) \; (2n+m) \; {\cal D}(2n,m) \; .
\end{equation}
The solution to Eq. (\ref{Eq:first_ex}), for $n=2$,
Eq. (\ref{Eq:second_ex}), for $n=4$, and this equation, is
the general result of Eq. (\ref{final}).  

For the case of interest, $m=3$, we obtain an especially simple result,
\begin{equation}
{\cal D}(2n,3) = (2n)!! \; (2n+1)!! = (2n+1)!  \; .
\label{comb_two}
\end{equation}

With this symmetry factor in hand, from Eq. (\ref{LLd})
the two point function of loops 
\begin{equation}
\langle L(x) L(0) \rangle  =  
\sum_{n=0}^\infty   \frac{(2n+1)!}{((2n)!)^2}
\left( \frac{g^2}{4 \, T} \; \Delta \right)^{2n}   = 
\cosh  \left( z \right) + 
z \; 
\sinh \left( z  \right) \;\;\; , \;\;\;
z = \frac{g^2}{16\pi} \, \frac{1}{x \, T} \; .
\label{two_colors_corr}
\end{equation}
At large distances, $z\rightarrow 0$ or $x \rightarrow \infty$,
\begin{equation}
\langle L(x) L(0) \rangle  
\approx 1 + \frac{3}{2} \; z^2 + \frac{5}{24} \; z^4 + \ldots
\approx 1 + \frac{3}{32} \left(\frac{g^2}{4 \pi}\right)^2 
\frac{1}{(x T)^2} + 
 \frac{5}{24} \left(\frac{g^2}{16 \pi x T}\right)^4
+ \ldots
\; .
\label{loop_two_large}
\end{equation}
For general $N_c$, to $\sim g^4$ 
\cite{Burnier:2009bk, Brambilla:2010xn}
\begin{equation}
\langle L^*(x) L(0) \rangle  
\approx 1 + \frac{N_c^2 - 1}{8 N_c^2} \left(\frac{g^2}{4 \pi}\right)^2 
\frac{1}{(x T)^2} + \ldots 
\;\;\; , \;\;\;  x \rightarrow \infty \; .
\label{loop_loop_large_distance_anyN}
\end{equation}
This result is due to the exchange of two gluons, with the factor
of $(N_c^2-1)/(2 N_c^2)$ from the Casimir for the fundamental
representation.  This agrees with Eq. (\ref{loop_two_large}) for
$N_c=2$.  Notice that the term $\sim z^4 \sim g^8$ is of higher
order than has been computed previously.  This is because we only
include tree diagrams.  

At short distances, $z \rightarrow \infty$, 
\begin{equation}
\langle L(x) L(0) \rangle  \approx
\frac{z}{2} \; \exp(z) \approx
\frac{1}{2} \exp\left( + \; \frac{g^2}{16 \pi} \; \frac{1}{x T} 
+ \log\left(\frac{g^2}{16 \pi x T}\right) \right) 
\;\;\; , \;\;\;  x \rightarrow 0 \; .
\label{short_dist_two}
\end{equation}
The coefficient in the exponential is the same as for
the Wilson loop at short distances, 
Eq. (\ref{Eq:WSU2F}) in Sec. (\ref{sec:Wilson_loop_2}).

What is novel here is that Eq. (\ref{two_colors_corr}) is valid for arbitrary
distances, and interpolates smoothly between small and large values
of $g^2/(x T)$.

\subsection{Analytic method}

We work with components, and use the multinomial theorem
\begin{equation}
\left(A_a(x)^2\right)^n = 
\sum_{k_1, k_2, k_3 = 1}^n  
\frac{n!}{k_1! \, k_2! \, k_3!} \;
A_1^{2 k_1}(x)  \; A_2^{2 k_2}(x)  \; A_3^{2 k_3}(x)  
\; \delta_{k_1+k_2+k_3,n} \; .
\end{equation}
After expanding each bracket in Eq.~\eqref{LLd} we ignore
self energy correction to match the powers
of $A^0$,
\begin{eqnarray}
\label{LLd2}
\langle L(x) && L(0) \rangle = \sum_{n=0}^\infty   \frac{1}{((2n)!)^2} 
\left( \frac{g}{2\, T} \right)^{4n} \\
&&  \times \sum_{k_1, k_2, k_3 = 0}^n 
 \left( \frac{n!}{k_1! k_2! k_3!}  \right)^2 
\left\langle 
A_1^{2 k_1}(x)  A_2^{2 k_2}(x)  A_3^{2 k_3}(x) 
A_1^{2 k_1}(0)  A_2^{2 k_2}(0)  A_3^{2 k_3}(0)  
\right\rangle
 \delta_{k_1+k_2+k_3,n} \; .
\nonumber
\end{eqnarray}
Since we have evaluated the fields component by component, it is
direct to evaluate the correlation functions,
\begin{eqnarray}
&&\left\langle 
A_1(x)^{2 k_1}  A_2(x)^{2 k_2}  A_3(x)^{2 k_3} 
A_1(0)^{2 k_1}  A_2(0)^{2 k_2}  A_3(0)^{2 k_3}  
\right\rangle
\nonumber
\\ 
&&= \left\langle 
A_1(x)^{2 k_1}  
A_1(0)^{2 k_1} 
\right\rangle
\left\langle 
A_2(x)^{2 k_2}  
A_2(0)^{2 k_2}  
\right\rangle
\left\langle 
A_3(x)^{2 k_3} 
A_3(0)^{2 k_3}  
\right\rangle 
\nonumber
\\ 
&&
= (2k_1)! \, (2k_2)! \, (2k_3)!  \, \Delta^{2k_1} \; \Delta^{2k_2} 
\, \Delta^{2k_3}  =  (2k_1)! \, (2k_2)! \, (2k_3)! \; \Delta^{2 n} \; .   
\end{eqnarray}
The last equality was obtained by taking into account the
relation $k_1+k_2+k_3=n$.

Next we represent the Kronecker symbol as
\begin{equation}
\delta_{k_1+k_2+k_3, n}  = \frac{1}{2\pi i} \oint_{C}
\frac{d\zeta}{\zeta} \; \zeta^{k_1+k_2+k_3-n} \; ,
\end{equation}
where the contour $C$ includes the origin.

The two point function becomes
\begin{equation}
\left\langle L(x) L(0) \right\rangle  = 
\frac{1}{2\pi i}\sum_{n=0}^\infty   
\frac{(n!)^2}{((2n)!)^2}  \left( \frac{g^2 }{4 \, T} \; \Delta\right)^{2n} 
\oint_C  \frac{d\zeta}{\zeta^{n+1}} 
\left( \sum_{k=0}^n  \frac{(2k)!}{(k!)^2} \zeta^k \right)^3  \; .
\end{equation}

We then evaluate the sum
\begin{equation}
\sum_{k=0}^n  \frac{(2k)!}{(k!)^2} 
\; \zeta^k = \frac{1}{\sqrt{1-4\, \zeta}} + \zeta^{n+1} F(\zeta) \; ,
\label{Eq:zeta1}
\end{equation}
where $F(\zeta)$ is holomorphic about the origin, $\zeta = 0$. 
Note the factor of $\zeta^{n+1}$ 
in front of $F(\zeta)$. Because of this, only the first term 
in Eq. (\ref{Eq:zeta1}) contributes,
\begin{equation}
\frac{1}{2\pi i}
\oint_C  \frac{d\zeta}{\zeta^{n+1}}  \left(  \frac{1}{\sqrt{1-4\, \zeta}}
+ \zeta^{n+1} F(\zeta) \right)^m   =  \frac{1}{2\pi i}
\oint_C  \frac{d\zeta}{\zeta^{n+1}} \; \frac{1}{(1-4\zeta)^{m/2}} \; .
\end{equation}
Since the function $F(\zeta)$ is complicated, this helps to 
simplify the problem.
For $m = 3$,
\begin{equation}
\frac{1}{\zeta^{n+1}} \; \frac{1}{(1-4\zeta)^{3/2} }
= \frac{1}{\zeta^{n+1}} \sum_{j=0}^\infty 2^{2j} \;
\frac{\Gamma(j + \frac{3}{2})}{j! 
\; \Gamma\left(\frac{3}{2}\right)} \;\zeta^{j} \; .
\end{equation}
Using the residue theorem,
\begin{equation}
\frac{1}{2\pi i}
\oint_C   \frac{d\zeta}{\zeta^{n+1}} 
\; \frac{1}{(1-4\zeta)^{3/2}} =   2^{2n} \; 
\frac{\Gamma \left(n + \frac{3}{2} \right))}{n! \, 
\Gamma \left(\frac{3}{2} \right)} \; .
\end{equation}
Altogether,
\begin{equation}
\langle L(x) L(0) \rangle  =  
\sum_{n=1}^\infty   \frac{n!}{((2n)!)^2} 
\frac{\Gamma\left(n+\frac{3}{2}\right)}{\Gamma\left(\frac{3}{2}\right)}
\left( \frac{g^2}{T} \Delta \right)^{2n}   \; .
\label{Eq:anal_corr}
\end{equation}
Using
\begin{equation}
\Gamma \left(n+\frac{3}{2} \right) 
= \frac{(2n+2)!}{4^{n+1} (n+1)!} \sqrt{\pi} \; ,
\end{equation}
it is direct to show that Eq. (\ref{Eq:anal_corr}) agrees with
the combinatoric result of Eq. (\ref{two_colors_corr}).

\subsection{Wilson loop}
\label{sec:Wilson_loop_2}

For completeness, in this section we compute a Wilson loop for two colors
in the tree approximation.  We take a Wilson loop which has the same
sides in imaginary time, and just tie the ends at $\tau = 0$ and
$\tau = 1/T$ together.  Since we drop self energy corrections, we can
drop the terms from $\tau = 0$ and $1/T$, and concentrate just on gluons
exchanged between the two sides, at $\vec{x}$ and $0$.

In general, there is no simple relation between
the exponential of the sum of two elements of the Lie algebra,
and the product of the corresponding exponentials.
In this case, however, path ordering implies that they do factorize:
\begin{equation}
 {\cal P} \exp\left( \frac{i \, g}{2\, T} 
\left(A_a(x) \sigma^a-A_b(0)\sigma^b \right) \right)
= \exp\left( \frac{i \, g}{2\, T} \; A_a(x) \sigma^a \right)
\exp\left( - \frac{i \, g}{2\, T} \; A_b(0) \sigma^b \right)
\label{path_ordering_factorize}
\end{equation}
There are several ways to establish this.  One can
check it to the first few orders in $g$, and then use an inductive proof
in powers of $A_a(x)$ and $A_b(0)$.  
A more elegant proof is the following.  Let
$A_a(x)$ be infinitesimally small, with $A_b(0)$ arbitrary.
Then by path ordering, $A_a(x)$ always lies to the left of any
$A_b(0)$.  For an infintesimal $A_a(x)$, to 
linear order in $A_a(x)$ we have
$\approx (1 + i g/(2 T) A_a(x) \sigma^a)
\exp( i g/(2 T) A_b(0) \sigma^b )$.  We can then write a differential
equation for $A_a(x)$, taking the derivative on the left.  The 
solution of this differential equation is the product of the 
exponentials in Eq. (\ref{path_ordering_factorize}).

Thus the Wilson loop is given by
\begin{eqnarray}
{\cal W}  = \left\langle\frac{1}{N_c} \; {\rm tr} \left[
\exp\left( \frac{ig}{2 T} \; A_a(x) \sigma^a\right) 
\exp\left( - \frac{ig}{2 T} \; A_b(0)\sigma^b \right) \right]
\right\rangle \; .
\label{wilson_loop}
\end{eqnarray}
We can now use the relation of Eq. (\ref{two_colors_expansion}),
\begin{eqnarray}
{\cal W}&=&
\left\langle\cos\left( \frac{g}{2 T} \sqrt{A_a(x)^2} \right) 
\cos\left( \frac{g}{2 T} \sqrt{A_b(0)^2} \right) 
\right\rangle
\nonumber \\
&+& 
\; 
\left\langle \; \widehat{A}_a(x) 
\widehat{A}_a(0)
\sin\left( \frac{g}{2 T} \sqrt{A_b^2(x)} \right) 
\sin\left( \frac{g}{2 T} \sqrt{A_c^2(0)} \right) 
\right\rangle\; .
\label{Eq:Wilson_loop_two}
\end{eqnarray}
For general $N_c$, the Wilson loop factorizes as above, but then
we cannot work out the expansion of the element of the Lie algebra.

The first term in Eq. (\ref{Eq:Wilson_loop_two}) is equal
to the correlator of two Polyakov loops.
The second term is new, but can be computed by similar means.
In a power series expansion, this second term equals
\begin{equation}
\label{Wilson_loop_sin}
\sum_{n=0}^\infty \; \frac{1}{((2n+1)!)^2} \; 
\left( \frac{g}{2 \, T} \right)^{4 n+2}
\left\langle  A_a(x) A_a(0) \left(A_b(x)^2\right)^n
\left( A_c(0)^2 \right)^n \right\rangle \; .
\end{equation}
As before, we neglect self energy corrections, so the powers of
$A_b(x)$ and $A_c(0)$ must match.

We need to solve a combinatoric problem very similar to that solved
previously, namely the number of ways to tie the fields
\begin{equation}
\left\langle  A_a(x) A_a(0) \left(A_b(x)^2\right)^n
\left( A_c(0)^2 \right)^n \right\rangle \; .
\end{equation}
together.  We denote this as ${\cal D}(2n+1,3)$.  It can computed
using our previous results for ${\cal D}(2n,3)$.
When $n=0$, we have just a sum over
one index $a$, so ${\cal D}(1,3) = 3$.

Now consider general $n$.  There are two cases
to consider.  For the first case, $A_a(x)$ can tie to $A_a(0)$, 
\begin{equation}
\left\langle  A_a(x) A_a(0) \right\rangle
\left\langle \left(A_b(x)^2\right)^n
\left( A_c(0)^2 \right)^n \right\rangle \; .
\end{equation}
This leaves the number of ways to tie
$\langle (A_b(x)^2)^n ( A_c(0)^2) ^n \rangle$ together.
Because of the sum over the index ``$a$'', there are
$3 \, {\cal D}(2n, 3)$ ways of tying the fields together.

The second case is if $A_a(x)$ ties to one of the other $A_c(0)$'s.  There
are $2 n$ such $A_c(0)$'s.  Similarly, $A_a(0)$ can then tie to
one of the $A_b(x)$'s in $2 n$ ways, leaving
\begin{equation}
\left\langle A_a(x) A_c(0) \right\rangle
\left\langle A_a(0) A_b(x) \right\rangle
\left\langle A_b(x) A_c(0) 
(A_d(x)^2)^{n-1}(A_e(0)^2)^{n-1} \right\rangle
\end{equation}
The last correlation function is related to ${\cal D}(2n-1,3$, so
in total there are $(2n)^2 \, {\cal D}(2n-1,3)$ ways to tying the fields
together.  

Thus we obtain 
\begin{equation}
{\cal D}(2n+1,3)
= 3 \, {\cal D}(2n,3)  +  (2n)^2 \, {\cal D}(2n-1,3) \; =
3 \, (2n+1)!  +  (2n)^2 \, {\cal D}(2n-1,3)
\label{inductive_odd}
\end{equation}
The solution to this equation is
\begin{equation}
{\cal D}(2n+1,3) = (2n+3) (2n+1)! \; .
\end{equation}
This satisfies ${\cal D}(1,3) = 3$, and one can check directly that
${\cal D}(3,3) = 30$.

The result is then
\begin{equation}
{\cal W} 
= (1 + z) \cosh(z) + (2+z) \sinh(z) \;\;\; , \;\;\;
z = \frac{g^2}{16\pi} \, \frac{1}{x \, T} \; .
\label{Eq:WSU2F}
\end{equation}

At large distances, $z \rightarrow 0$, and
\begin{equation}
{\cal W} \approx 1 + 3 \, z + \ldots \approx 1 + 
\frac{3 g^2}{16 \pi} \; \frac{1}{x T} + \ldots
\;\;\; , \;\;\; x \rightarrow \infty \; .
\end{equation}
This is the expected Coulomb term from single gluon exchange.

At short distances, the exponential agrees with the result from the
line-line correlator, Eq. (\ref{short_dist_two}).  
\begin{equation}
{\cal W} \approx z \, \exp(z) \approx \exp
\left(
+ \; \frac{g^2}{16 \pi} \; \frac{1}{x T} 
+ \log\left(\frac{g^2}{16 \pi x T}\right) \right)  + \ldots
\;\;\; , \;\;\; x \rightarrow 0 \; .
\end{equation}
The Wilson loop is similar to 
the exponential of the Coulomb term, but the coefficient
of $\sim 1/(x T)$ differs by a factor of three.  It is for this
reason that we refer to the behavior at short distances as Coulomb-like.
Notice also that in the exponent, there is a factor of $\log(z)
\sim \log(g^2/(xT))$, which has no analogy in perturbation theory.
We discuss this further in Sec. (\ref{sec:Conclusions}).

\section{Large $N_c$}

For two colors the only difficult part of the problem was in taking into
account all of the combinatoric factors from contracting powers of
$\sum_{a}(A_a)^2$ together.  For more than two colors, there
are two difficulties.

The first is the same problem, keeping track of combinatoric powers of
$\sum_{a} (A_a)^2$, where the sum over the color index ``$a$'' runs from
$1$ to $N_c^2-1$.  
In this section we show that in the
planar limit, taking $N_c \rightarrow \infty$ while holding $g^2 N_c$
fixed, that the combinatorics is {\it much} simpler
even than two colors.  

This simplification is special to the limit of large $N_c$. 
The other problem is that for finite $N_c > 2$, the symmetric
structure constant, $d^{a b c}$, also arises in the expansion of the
exponential.  For three or more colors, there is no 
general expression which relates an element of the Lie algebra, in this
case $A^0_a$, to one in the Lie group.  For reasons which are
clear after the fact, this complication can be ignored in the planar limit.

In Appendix (\ref{sec:pert_anyN}) and (\ref{sec:appendix}),
we present the results of computations which interpolate
between $N_c = 2$ and $\infty$, for some of the lowest order coefficients
which arise.  They demonstrate the correctness of our results for
$N_c=2$ and $\infty$, but show that the case of general $N_c$ is
not elementary.

While we can compute in the planar limit, we do not find Coulomb-like
behavior at short distances.  
Nevertheless, we present the results of our computation, since they are
relatively easy to obtain.  

For three or more colors, the loop and anti-loop are not equal.  We begin
with the case of the correlator between a loop and an anti-loop, as that
is more transparent.  Order by order in perturbation theory,
\begin{equation}
\langle L^*(x) L(x) \rangle
= \sum_{n=1}^{\infty} \;
\frac{1}{(n!)^2} \left(\frac{g}{T}\right)^{2n}
\left\langle \Tr{A^n(x)} 
\Tr{  A^n(0) } \right\rangle  
\label{loop_antiloop}
\end{equation}
As for two colors we neglect self energy corrections, so the
number of $A(x)$'s and $A(0)$'s must match.  The first term,
from $n=1$, vanishes because of the color trace.

For arbitary $N_c$, the problem is computing the color traces
in the above expression.
Let us then consider the possible contractions of powers of $A(x)$ with
those of $A(0)$.  We can always choose the first $A(x)$ to be left most
in the trace.  There are then ``$n$'' ways to connect this $A(x)$ with
any of the $A(0)$'s.  We can then make this color matrix the left most
in its trace as well,
\begin{equation}
{\cal D}_n = \left\langle \Tr{A^n(x)} 
\Tr{  A^n(0) } \right\rangle  
=  n \; \left( \frac{\Delta}{T} \right)   
\left\langle \Tr{t^a A^{n-1}(x)} 
\Tr{  t^a A^{n-1}(0) } \right \rangle  \; .
\label{trn0} 
\end{equation}

To understand our results, consider first
the case of an Abelian theory.  This is much simpler,
because then there are no color traces to
bother with.  For the first pair of photons,
the factor of ``$n$'' is the same as above.
For the second pair of photons, there are $n-1$ ways to tie the second
photon $A(x)$ to one of the other
$n-1$ $A(0)$'s.  This continues, so that in all
there are $n!$ different ways of connecting the photons together.  This cancels
one of the factors of $1/n!$ in Eq. (\ref{loop_antiloop}), giving
Eq. (\ref{abelian}).  In the present case, this is
\begin{equation}
\exp\left( + \; \frac{e^2}{4 \pi} \; \frac{1}{x \, T} \right)
\end{equation}
This result, from single photon exchange, is valid at all distances.
At large distances, this equals
$\approx 1 + e^2/(4 \pi x T) + \ldots$, which is precisely
the Coulomb term.  That is, in the Abelian theory the Coulomb term
directly exponentiates.  

In the planar limit, however, the combinatoric factors are trivial to
compute.  We first give an elementary argument, and then a detailed
analysis to verify that.

As a check, in Sec. (\ref{sec:pert_anyN}) we have computed the results
at finite $N_c$ for the first twelve coefficients of the perturbative
expansion of the loop anti-loop correlation functions.  The results
for the first six terms agree with previous results by 
Burnier, Laine, and Vepsalainen \cite{Burnier:2009bk}
and by Brambilla, Ghiglieri, Petreczky, and Vairo
\cite{Brambilla:2010xn}.  Note that we only compare the diagrams at
tree level, which are much easier than the computations to higher loop
order performed by these authors.

\begin{figure}[t]
\begin{center}
\includegraphics[width=0.15\textwidth]{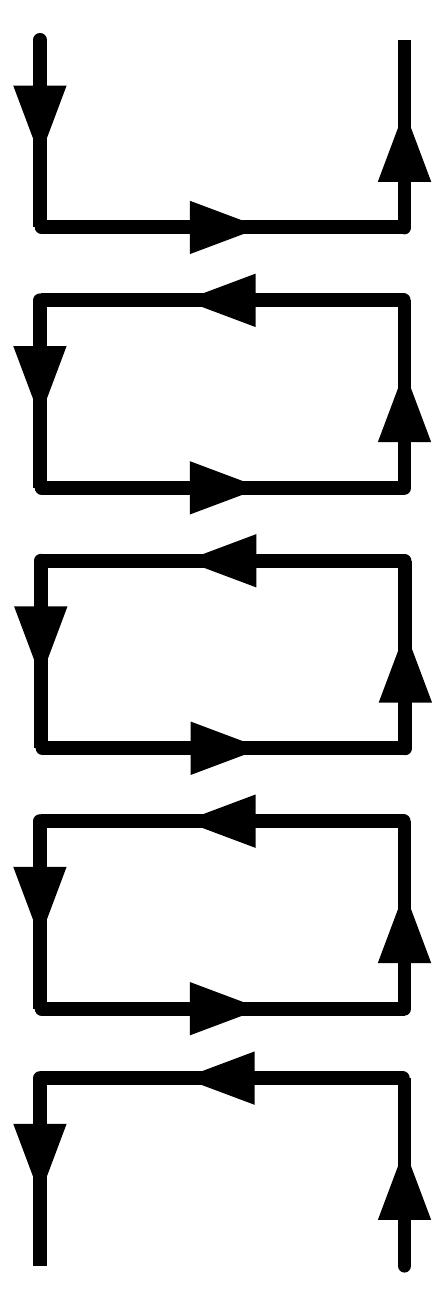}
\caption{
The planar diagram that dominates at infinite $N_c$ for the loop
anti-loop correlation function, in
the double line notation.  Notice that since we use static gauge,
that the ordering of the lines is directly in color space.
}
\label{fig:planar}
\end{center}
\end{figure}

Start with Eq. (\ref{trn0}), where we have connected one $A(x)$ to one
$A(0)$.  Consider first attaching the second gluon $A(x)$, which is
immediately to the right of $t^a$ in the color trace.
Using the double line notation, this gives the planar diagram
of Fig. (\ref{fig:planar}).

Suppose, instead, that we had attached the second gluon $A(x)$ to the 
{\it third}
$A(0)$.  This diagram is not planar, and so suppressed by $1/N_c^2$.
Indeed, it is clear diagramatically that the planar diagrams are
strictly ordered in color: the second $A(x)$ connects to the second
$A(0)$, the third $A(x)$ to the third $A(0)$, and so on.

Thus in the planar limit, the combinatoric factor is simply $n$.
This is in contrast to the Abelian theory, where 
the corresponding factor is $n!$, and
the case of two colors, where the factor is $(n+1)!$.

We now proceed to an equivalent argument, which is useful in computing
corrections in $1/N_c^2$ to the planar limit.
We start with the identity
\begin{equation}
{\rm tr} \left[ t^a M_1\right] \left[ t^a M_2   \right] = 
\frac{1}{2} {\rm tr}\left[  M_1   M_2\right]
- \frac{1}{2N_c}  \; {\rm tr}   M_1  \; {\rm tr} M_2 \; .
\end{equation}
to reduce
\begin{eqnarray}
D_n = n \; \frac{\Delta}{2 T} \; 
\left\langle  \Tr{A^{n-1}(x)A^{n-1}(0)}  - \frac{1}{N_c} 
\Tr{A^{n-1}(x)} \Tr{A^{n-1}(0)}\right\rangle \; .
\label{trn}
\end{eqnarray}
It is not evident, but one can show that the connected
part of the second term is suppressed in the limit of large $N_c$.

Thus we concentrate on the first term.  Note that 
terms of the highest order in $N_c$ originate from contractions as follows.
Considering the term with three gluons, this is, graphically,
\begin{equation}
\contraction[1ex]{}{A(x)}{A(x)A{x}A(0)
A(0)}{A(0)}
\contraction[3ex]{A(x)}{A(x)}{A(x) A(0)}{A(0)}
\contraction[4ex]{A(x)A(x)}{A(x)}{}{A(0)}
A(x)A(x)A(x)A(0)A(0)A(0)
\label{contractions}
\end{equation}
Note that $t^a t^a = C_f \mathbb{I}\approx N_c/2 \; \mathbb{I}$
at large $N_c$.  
Using this, we work gluon by gluon to reduce the trace,
\begin{eqnarray}
&&
\left\langle 
\Tr{A^{a_1}(x) t^{a_1} A^{a_2}(x) t^{a_2} 
A^{a_3}(x) t^{a_3}  A^{b_1}(0) t^{b_1}  A^{b_2}(0) t^{b_2}  
A^{b_3}(0) t^{b_3}   }
\right\rangle
\approx \\ 
\nonumber 
&&\Delta(x) 
\left\langle
\Tr{ t^{a_1}  A^{a_2}(x) t^{a_2}  A^{a_3}(x) t^{a_3}  
A^{b_1}(0) t^{b_1}  A^{b_2}(0) t^{b_2}  t^{a_1}   }
\right\rangle
= \\
&&
\Delta(x) C_f 
\left\langle
\Tr{ A^{a_2}(x) t^{a_2}  A^{a_3}(x) t^{a_3}  
A^{b_1}(0) t^{b_1}  A^{b_2}(0) t^{b_2}   } 
\right\rangle
\approx 
\\ &&\Delta^2(x) C_f 
\left\langle 
\Tr{ t^{a_2}    A^{a_3}(x) t^{a_3}  A^{b_1}(r) t^{b_1} t^{a_2}  } 
\right\rangle
= \\
&& \Delta^2(x) C_f^2 
\left\langle 
\Tr{    A^{a_3}(x) t^{a_3}
A^{b_1}(x) t^{b_1} }
\right\rangle
\approx
\Delta^3(x) C_f^3 \Tr{\mathbb{I}} = N_c \Delta^3(x) C_f^3. 
\label{proof3}
\end{eqnarray}
The final factor of $N_c$ comes from the trace of a unit 
matrix after all contractions are done.

In general, 
\begin{eqnarray}
&& D_n(x) \equiv \left\langle  
\Tr{A^n(x)} \Tr{  A^n(0) }  \right\rangle  
\underset{N_c\to \infty}=  \frac{n \Delta(x)}{2} N_c 
\left(\frac{N_c \Delta(x)}{2} \right)^{n-1} = 
n \left(\frac{N_c \Delta(x)}{2} \right)^{n} \; .
\label{trn1}
\end{eqnarray}
The sum of the power series gives a Bessel Function $I_1$,
\begin{eqnarray}
\langle L^*(x) L(0) \rangle  
&=& 1 + \frac{1}{N_c^2} \sum_{n=2}^{\infty} 
\frac1{n! (n-1)!}  \left(\frac{g^2 N_c }{2 \, T} \; \Delta(x)\right)^{n} 
\nonumber
\\
&=&
1 + \frac{1}{N_c^2} \left(  z^{1/2} I_1(2 z^{1/2}) - z \right) \;\;\; , \;\;\;
z= \frac{g^2 N_c}{8 \pi} \, \frac{1}{x T} \; .
\label{loop_antiloop_result} 
\end{eqnarray}
This is our final result for the loop anti-loop correlator.

Similar arguments can be used to compute the correlator between two
Polyakov loops.  This is the correlator between a (infinitely heavy)
test quark and another test quark.  At first the color structure in
the planar limit looks different, but one can recognize the following.
Take the planar diagram for the loop anti-loop correlator, and flip the 
line for the anti-loop on its head.  This is then the planar diagram for
the loop loop correlator.  

The only change is then the sign of the charges.  For the loop anti-loop
correlation function, each term in Eqs. 
(\ref{loop_antiloop}) and (\ref{loop_antiloop_result}) are positive,
as the loop has charge $+ ig $, and the anti-loop, charge $- i g$.
For the loop loop correlation function, each has charge $+ ig$, so there
is a factor of $(-1)^n$ in the sum,
\begin{eqnarray}
\langle L(x) L(0) \rangle  
&=& 1 + \frac{1}{N_c^2} \sum_{n=2}^{\infty} 
\frac{(-1)^n}{n! (n-1)!}  
\left(\frac{g^2 N_c }{2 \, T} \; \Delta(x)\right)^{n} 
\nonumber
\\
&=&
1 + \frac{1}{N_c^2} \left(-  z^{1/2} J_1(2 z^{1/2}) + z \right) \;\;\; , \;\;\;
z= \frac{g^2 N_c}{8 \pi} \, \frac{1}{x T} \; .
\label{loop_loop}
\end{eqnarray}

The propagator is a function only of the absolute value of the spatial
separation, $x$.  Then for the correlation functions, only the relative
charges matter.  Thus the correlation function between a loop and an
anti-loop, with charges $ig$ and $-ig$, is the same as 
between an anti-loop and a loop, with charges $-ig$ and $ig$,
\begin{equation}
\langle L^*(x) L(0) \rangle  =
\langle L(x) L^*(0) \rangle  \; .
\end{equation}
Similarly, the correlation function between two loops, each
with charge $ig$, equals that between two anti-loops, each with charge
$-ig$,
\begin{equation}
\langle L(x) L(0) \rangle  =
\langle L^*(x) L^*(0) \rangle  \; .
\end{equation}

We can now decompose the loop into real and imaginary parts.  The above
identities imply that there is no correlation between the real and
imaginary parts, only between themselves.  That for the real part is
\begin{equation}
\langle {\rm Re} L(x) \; {\rm Re} L(0)  \rangle  = 1 +  
\frac{z^{1/2}}{2N_c^2} \left[ I_1(2 z^{1/2}) - J_1(2 z^{1/2})  \right] 
\end{equation}
and those for the imaginary parts,
\begin{equation}
\langle {\rm Im} L(x) \; {\rm Im} L(0)  \rangle  = 
\frac{1}{N_c^2} 
\left(-z+\frac{z^{1/2}}{2} \left[ I_1(2 z^{1/2}) + J_1(2 z^{1/2})  \right] 
\right)
\end{equation}

The Wilson loop can also be computed directly in the planar limit.  By
path ordering, we put all powers of $A(x)$ to the left, and all those
of $A(0)$ to the right.  Neglecting self energy corrections,
we expand to ``$n$'' powers of $i g A(x)$, and the same for $- i g A(0)$.
There is a combinatorial factor of $1/n!$ for each term, $\sim A(x)^n$
and $\sim A(0)^n$.
A planar diagram is given by connecting the 
left most $A(x)$ to the left most $A(0)$.  Doing so, we then have to
connect the second most left $A(x)$ to the second most left $A(0)$.
Continuing in this way, there is no extra factor of ``$n$'', and
the Wilson loop is given by
\begin{equation}
{\cal W} = \frac{1}{N_c}\sum_{n=0}^\infty 
\left(\frac{g}{T}\right)^{2 n} \frac{1}{(n!)^2} 
\left\langle  \Tr{ A^n(x) A^n(0) } \right\rangle = 
\sum_{n=0}^\infty 
\frac{1}{(n!)^2} 
\left(\frac{g^2 N_c }{2 T}\; \Delta(x)\right)^n =
I_0\left( 2 z^{1/2} \right)  
\label{Eq:WSUinfty}
\end{equation}

At large distances, $z \rightarrow 0$, the correlation function for
the real part of the loop is
\begin{equation}
\langle {\rm Re} L(x) {\rm Re} L(0) \rangle  \approx
1 + \frac{1}{2 N_c^2}\left(
z^2 + \frac{z^4}{72} \ldots \right)
=
1 + \frac{1}{8 N_c^2} \left( \frac{g^2 N_c}{4 \pi} \right)^2 \;
\left(\frac{1}{x \,T} \right)^2 + \ldots \, 
\end{equation}
The first term, $=1$, is from the disconnected graph, which dominates
at large $N_c$.  The second term is due to connected graphs,
which are down by $1/N_c^2$.  
The coefficient of the term to leading order,
$\sim g^2 N_c$, agrees with the large $N_c$ limit of Eq.
(\ref{loop_loop_large_distance_anyN}).  

For the correlators of the imaginary part of the loop, there are
no contributions from disconnected diagrams, only from connected 
diagrams, which are then necessarily $\sim 1/N_c^2$.  For large
distances, $z \rightarrow 0$, it begins at $\sim z^3$,
which is $g^6$:
\begin{equation}
\langle {\rm Im} L(x) {\rm Im} L(0) \rangle  \approx
\frac{1}{N_c^2}\left(
\frac{z^3}{12} + \frac{z^5}{2880} \ldots \right)
=
\frac{1}{96 N_c^2} \left( \frac{g^2 N_c}{4 \pi} \right)^3 \;
\left(\frac{1}{x \,T} \right)^3 + \ldots \, 
\label{im_part_large}
\end{equation}

For the Wilson loop, at large distances it behaves as
\begin{equation}
{\cal W} \approx 
1 + z + \frac{z^2}{4} + \frac{z^3}{36} + \ldots 
\approx 1 + \frac{g^2 N_c}{8 \pi} \; \frac{1}{x\, T} + \ldots \; .
\end{equation}
The expectation value of the Wilson loop is in and of itself a
disconnected diagram, and so all contributions survive in the planar
limit at infinite $N_c$.  The leading term, $\sim g^2$, is from
single gluon exchange.

What is less obvious is the limit of short distances.  
For small $z$, 
\begin{equation}
I_{\alpha}(z) \to  \frac{e^z}{\sqrt{2\pi z}} \;\;\; , \;\;\; \quad 
J_1(z) \to  {\frac {\sin \left( z \right) -
\cos \left( z \right) }{\sqrt {\pi z }}} \; ,
\end{equation}
we find that the behavior of {\it all} correlation functions at
small distances is not the exponential of
$\sim 1/r$, but the exponential of $\sim 1/\sqrt{r}$.
For example, for the Wilson loop
\begin{equation}
{\cal W} \approx \exp\left(2 \sqrt{z}\right)
\approx \exp\left(2 \sqrt{\frac{g^2 N_c}{8 \pi}} \; 
\sqrt{\frac{1}{ x T}} \right)
\; .
\label{Wilson_loop_infN_short}
\end{equation}

The behavior at short distances can also be derived directly from
the sum over $n$.  At short distances, $z$ is large, and one
can show that the dominant behavior is given by asymptotically large
$n$.  To show this, one can use an effective action.  For the
Wilson line at large $N_c$, it is 
\begin{equation}
\frac{1}{(n!)^2} \; z^n
= \exp\left( {\cal S}_{eff}(n) \right) \; ,
\end{equation}
where
\begin{equation}
{\cal S}_{eff} (n) = - 2 \log(n!) + n \log z
\approx_{n \rightarrow \infty}
- 2 \left( n \log(n) - n \right) + n \log z \; .
\end{equation}
At large $n$, the stationary point of the effective action
$\partial {\cal S}_{eff}/\partial n = 0$, is given by
$n_0 = \sqrt{z}$.  Then ${\cal S}_{eff}(n_0) \sim 
2 n_0 = 2 \sqrt{z}$, in accord with 
Eq. (\ref{Wilson_loop_infN_short}).  The same method can be used
to derive the short distance behavior of the other correlation functions.
This is perfectly reasonable: the behavior at large distances, when
$z$ is small, is determined by the first few orders in perturbation theory.
That at short distances, when $z$ is large, is determined by
the asymptotically large orders of perturbation theory.

\section{Conclusions}
\label{sec:Conclusions}

In this paper we computed the correlators
of Polyakov (and Wilson) loops at tree level, as functions of
$g^2 N_c/(x T)$.  In this section we suggest that the 
non-Coulombic behavior at short distances in Eq.
(\ref{Wilson_loop_infN_short}) is an artifact of the
planar limit at tree level.

We first review, on a heuristic level, why one expects Coulombic
behavior at short distances.
As we work in imaginary time, the length in the time direction is
$\beta = 1/T$.  Now consider the limit of $\beta \rightarrow \infty$.
The discussion is similar for either
the Wilson loop or the correlation function between the loop and anti-loop
(or loop loop).  For simplicity we discuss the Wilson loop, since all
contributions persist in the planar limit.

Single gluon exchange gives a contribution $\sim g^2 N_c \beta/x$.  
Two gluon exchange gives rise to two terms.  The two gluons can be
moved independently up and down the time axis, which gives
the square of the first term, $\sim (g^2 N_c \beta/x)^2$.  There
are also terms where the two gluons do not move independently,
$\sim (g^2 N_c)^2 \beta/x$.  
The latter represents corrections at one loop order to the potential
at tree level.

For our purposes we can ignore perturbative corrections to the
tree level potential.  The leading term is
from the exchange of ``$n$'' gluons, which move independently up and down
the time axis.  These one gluon kernels are identical, and so there is
an associated combinatoric factor of $1/n!$.  This then generates
the exponential of the potential at tree level.
Because we assume that $\beta \rightarrow \infty$ at the outset,
by dimensional analysis the coefficient of any term $\sim \beta$ is
necessarily Coulombic, $\sim 1/x$.  
This heuristic analysis underlies extensive studies in perturbation
theory, beginning with Ref. 
\cite{Fischler:1977yf, *Appelquist:1977tw, *Appelquist:1977es}.

A more modern approach uses effective theories, such as
potential non-relativistic QCD \cite{Brambilla:2010xn}.
For the Wilson loop, or the loop anti-loop correlator, at short distances
one relates these correlation functions to two fields, for a color
singlet and a color octet.   The effective theory
approach directly allows the exponentiation of the potential in each
channel, as the two point function is equal to the exponential of the
potential, times the expectation value of the associated Polyakov
loop in either the singlet or octet representations.

The same analysis can be carried out for the loop loop correlator.
The operators which enter are those for operators with two indices
in the fundamental representation.  There are two irreducible representations,
either symmetric or anti-symmetric in the two indices.
(For example, for three colors these are the
sextet and anti-triplet representations, respectively.)  
For each representation,
the two point function of the loop loop correlator is
then related to the exponential of the 
associated potential, 
times the expectation value of the Polyakov loop in the
associated representation.  

This analysis using effective theory applies in weak coupling for
arbitrary $N_c$, in the limit that $\beta \rightarrow \infty$.
This is different from our computations, where the parameter
$g^2 N_c \beta/x$ is kept finite throughout.  
Consequently, we believe that our result is peculiar to the planar
approximation.  We do not understand why the result at short
distances is weaker than Coulombic in the planar limit.

We conclude by discussing the implications of our results for
the work of 
Lo, Friman, Redlich, and Sasaki \cite{Lo:2013etb,*Lo:2013hla}.  
Usually what is computed in non-Abelian gauge theories are the expectation
values of (renormalized) Polyakov loops in different representations.
The process of renormalization is not simple on the lattice
\cite{Kaczmarek:2002mc, Dumitru:2003hp, Gupta:2007ax, Mykkanen:2012ri}.  While
the self energy corrections to the loop vanish with dimensional regularization,
they do not on the lattice.  Instead, they give rise to
terms $\sim \exp(- C_{\cal R} g^2/(a T))$, 
where ``$a$'' is the lattice spacing,
and $C_{\cal R}$ the Casimir for the loop in an irreducible representation
${\cal R}$.  By the character expansion, products of loops at the same
point reduce to a linear sum of loops in irreducible representations
\cite{Dumitru:2003hp}.  

In Refs. \cite{Lo:2013etb,*Lo:2013hla} 
Lo, Friman, Kaczmarek, Redlich, and Sasaki study the susceptibility
of Polyakov loops for three colors.  What is especially interesting is
the difference between the susceptibilities for the real and imaginary
parts of the loop.  

For local fields, there is no concern about the measurement
of such susceptibilities.  If one measures $\phi(x) \phi(0)$, then
by the Operator Product Expansion there are new ultraviolet divergences
as $x \rightarrow 0$.  However, these are local terms.  A susceptibility
involves the average of each field over the volume of space (or space-time),
$\cal V$.  Thus while new divergences will arise, accompanied
by powers of, {\it e.g.}, the lattice spacing ``$a$'', in the end
they will be overwhelmed by an overall factor of $1/{\cal V}$.  That is,
the divergences associated with bringing two operators together does
not affect the measurement of the susceptibility.

Polyakov loops, however, are non-local operators, and on the lattice,
have exponential divergences.  Thus ultraviolet divergences, which
are proportional to $\sim \exp(+ \#/(a T)^m)$, will overwhelm factors
of the volume, $1/{\cal V}$, in the continuum limit, as $aT \rightarrow 0$.
For operators which are dominated by Coulomb behavior at small distances,
$m=1$.  In the planar limit we find $m= 1/2$, but we believe this is
special to large $N_c$.  For the imaginary part of the loop, the leading
perturbative correction is from three gluon exchange, 
Eq. (\ref{im_part_large}), and so then $m=3$.  It appears necessary to
eliminate the effects of these
divergences before one can extract susceptibilities
of the Polyakov loops which are free from effects of the lattice cut-off.

\begin{acknowledgements}
We thank F. Karsch and S. Mukherjee for discussions about
Ref. \cite{Lo:2013etb,*Lo:2013hla} which
inspired our interest in this problem.  We also thank
L. McLerran and P. Petreczky for useful discussions.
R.D.P.  is supported
by the U.S. Department of Energy under contract \#DE-AC02-98CH10886.
\end{acknowledgements}

\appendix

\section{Perturbative expansion at arbitrary $N_c$}
\label{sec:pert_anyN}

In the perturbative expansion of the correlation functions, we need
\begin{equation}
{\cal D}_n = \left\langle \Tr{A^n(x)} 
\Tr{  A^n(0) } \right\rangle  
= c_n \; \left(\frac{\Delta(x)}{T}\right)^n \; .
\end{equation}
By brute force, the first eleven coefficients are
\begin{eqnarray}
&&c_2=\frac{N_c^2}{2}-\frac{1}{2} \; , \\ 
&&c_3=\frac{3 N_c^3}{8}-\frac{15 N_c}{8}+\frac{3}{2 N_c} \; , \\
&&c_4=\frac{N_c^4}{4}-\frac{7 N_c^2}{4}+6-\frac{9}{2 N_c^2} \; , \\
&&c_5=\frac{5 N_c^5}{32}-\frac{25 N_c^3}{32}+\frac{35 N_c}{8}
-\frac{75}{4 N_c}+\frac{15}{N_c^3} \; , \\ 
&&c_6=\frac{3 N_c^6}{32}+\frac{15 N_c^4}{32}
-\frac{99 N_c^2}{16}+\frac{45}{8} +
\frac{225}{4 N_c^2}-\frac{225}{4 N_c^4} 
\; , \\
&&c_7=\frac{7 N_c^7}{128}+\frac{49 N_c^5}{32}-\frac{2597 N_c^3}{128}
+\frac{6657 N_c}{64}-\frac{735}{4 N_c}-\frac{2205}{16 N_c^3}
+\frac{945}{4 N_c^5} \; , \\ 
&&c_8=\frac{N_c^8}{32}+\frac{35 N_c^6}{16}-\frac{875 N_c^4}{32}
+\frac{1545 N_c^2}{8}-903+\frac{3675}{2 N_c^2}-\frac{2205}{2 N_c^6} \; , \\ 
&&c_9=\frac{9 N_c^9}{512}+\frac{621 N_c^7}{256}-\frac{9639 N_c^5}{512}
+\frac{9873 N_c^3}{128}-\frac{11745 N_c}{16}+\frac{90153}{16 N_c}
-\frac{59535}{4 N_c^3}\nonumber \\
&&\;\;\;\;\;\;\;\;\;\;
+\frac{8505}{2 N_c^5}+\frac{5670}{N_c^7} \; , \\
&&c_{10}=\frac{5 N_c^{10}}{512}+\frac{75 N_c^8}{32}
+\frac{3705 N_c^6}{512}-\frac{94975 N_c^4}{256}+\frac{95265 N_c^2}{32}
-\frac{31725}{4}-\frac{155925}{8 N_c^2}
\nonumber\\
&& \;\;\;\;\;\;\;\;\;\;
+\frac{439425}{4 N_c^4}-\frac{212625}{4 N_c^6}-\frac{127575}{4N_c^8}
 \; , \\ 
&&c_{11}=
\frac{11 N_c^{11}}{2048}
+\frac{4235 N_c^9}{2048}
+\frac{94743 N_c^7}{2048}
-\frac{2051555 N_c^5}{2048 }
+\frac{9832823 N_c^3}{1024}
-\frac{15454395 N_c}{256}
\nonumber\\ 
&&\;\;\;\;\;\;\;\;\;\;
+\frac{13476375}{64 N_c} 
-\frac{4459455}{32N_c^3}
-\frac{12006225}{16 N_c^5} 
+\frac{8575875}{16 N_c^7}
+\frac{779625}{4 N_c^9}
\; , \\
&&c_{12}=
\frac{3 N_c^{12}}{1024}
+\frac{1749 N_c^{10}}{1024}
+\frac{92169 N_c^8}{1024}
-\frac{1365573 N_c^6}{1024}
+\frac{2542947 N_c^4}{256}
-\frac{9823527 N_c^2}{128}
 \nonumber\\
&&\;\;\;\;\;\;\;\;\;\;
+\frac{39051045}{64}
-\frac{45721665}{16 N_c^2}
+\frac{33960465}{8 N_c^4}
+\frac{36018675}{8 N_c^6}
-\frac{5145525}{N_c^8}
-\frac{5145525}{4 N_c^{10}}\; .
\end{eqnarray}

For two colors, all $c_n$ vanish for odd $n$, while for even $n$,
\begin{equation}
c_n = (n+1)! \; \frac{1}{2^{2n-2}} \; ,
\end{equation}
which agrees with our previous analysis: the factor of $(n+1)!$
is ${\cal N}(n,3)$ in Eq. (\ref{comb_two}).

\section{Corrections at large $N_c$}
\label{sec:appendix}

From the above results, we could not deduce the general result for arbitrary
$n$ and $N_c$.  However, at large $N_c$ the first two
terms satisfy
\begin{equation}
c_n \approx n \left(\frac{N_c}{2}\right)^n 
\left( 1 + \frac{1}{24 \, N_c^2} \; n(n-1)(n(n-1) - 26) 
+ O\left(\frac{1}{N_c^4} \right)\right)
\end{equation}
The first term, $= n (N_c/2)^n$, agrees with our combinatorial analysis
above.  The 
second term is admittedly a guess, satisfied by $c_2 \ldots c_{12}$.

We now perform the exercise to assume that this guess is
correct, and use it compute the leading correction in $1/N_c^2$.  

The leading order in $1/N_c^2$ to the correlator for the loop
anti-loop correlation function is
\begin{equation}
\frac{1}{N_c^4} \left(\frac78 z^2 + \frac{1}{12} z^2 I_0(2\sqrt{z}) + 
\frac{z^{3/2}}{24} (z-26) I_1(2\sqrt{z})\right) \; .
\label{Eq:corrlal}
\end{equation}

For the loop loop correlator, one changes $z\to -z$,
\begin{equation}
\frac{1}{N_c^4} \left(\frac78 z^2 + \frac{1}{12} z^2 J_0(2\sqrt{z}) 
- \frac{z^{3/2}}{24} (z+26) J_1(2\sqrt{z})\right) \; .
\label{Eq:corrll}
\end{equation}

Guessing the corrections to higher order is not trivial.
Those $\sim 1/N_c^4$ are proportional to
\begin{equation}
\frac{9 }{16}n^6 -\frac{4567 }{240}n^5
+\frac{4289 }{16}n^4 -\frac{96733}{48} n^3
+\frac{67251 }{8}n^2 -\frac{179387 }{10}n +14553 \; .
\label{Eq:NNLO}
\end{equation}
when $n > 4$.

\bibliography{loop}

\end{document}